\documentclass[10pt, conference, compsocconf]{IEEEtran}

\usepackage{url}
\usepackage{cite}
\usepackage{amsmath,amssymb,amsfonts}
\usepackage{algorithmic}
\usepackage{graphicx}
\usepackage{textcomp}
\usepackage{xcolor}
\usepackage{subcaption}
\usepackage{listings}
\usepackage{xcolor}
\usepackage{booktabs}

\definecolor{background}{RGB}{250,250,250}
\definecolor{numb}{RGB}{220,0,0}
\definecolor{punct}{RGB}{0,0,0}
\definecolor{delim}{RGB}{0,0,255}

\lstdefinelanguage{json}{
    basicstyle=\scriptsize\ttfamily,
    numbers=none,
    numberstyle=\scriptsize,
    stepnumber=1,
    numbersep=8pt,
    showstringspaces=false,
    breaklines=true,
    frame=lines,
    backgroundcolor=\color{background},
    literate=
     *{0}{{{\color{numb}0}}}{1}
      {1}{{{\color{numb}1}}}{1}
      {2}{{{\color{numb}2}}}{1}
      {3}{{{\color{numb}3}}}{1}
      {4}{{{\color{numb}4}}}{1}
      {5}{{{\color{numb}5}}}{1}
      {6}{{{\color{numb}6}}}{1}
      {7}{{{\color{numb}7}}}{1}
      {8}{{{\color{numb}8}}}{1}
      {9}{{{\color{numb}9}}}{1}
      {:}{{{\color{punct}{:}}}}{1}
      {,}{{{\color{punct}{,}}}}{1}
      {\{}{{{\color{delim}{\{}}}}{1}
      {\}}{{{\color{delim}{\}}}}}{1}
      {[}{{{\color{delim}{[}}}}{1}
      {]}{{{\color{delim}{]}}}}{1},
}

\begin{document}
%
\title{FUNSD: A Dataset for Form Understanding \\ in Noisy Scanned Documents}


\author{\IEEEauthorblockN{Guillaume Jaume\footnote{Corresponding author}}
\IEEEauthorblockA{Swiss Federal Institute of Technology\\
Signal Processing Laboratory 5 \\ 
Lausanne, Switzerland \\
guillaume.jaume@epfl.ch}
\and 
\IEEEauthorblockN{Haz{\i}m Kemal Ekenel}
\IEEEauthorblockA{Istanbul Technical University\\
Department of Computer Engineering \\
Istanbul, Turkey \\
ekenel@itu.edu.tr}
\and
\IEEEauthorblockN{Jean-Philippe Thiran}
\IEEEauthorblockA{Swiss Federal Institute of Technology\\
Signal Processing Laboratory 5 \\ 
Lausanne, Switzerland \\
jean-philippe.thiran@epfl.ch}
}


\maketitle

\begin{abstract}
We present a new dataset for form understanding in noisy scanned documents (FUNSD) that aims at extracting and structuring the textual content of forms. The dataset comprises 199 real, fully annotated, scanned forms. The documents are noisy and vary widely in appearance, making form understanding (FoUn) a challenging task. The proposed dataset can be used for various tasks, including text detection, optical character recognition, spatial layout analysis, and entity labeling/linking. To the best of our knowledge, this is the first publicly available dataset with comprehensive annotations to address FoUn task. We also present a set of baselines and introduce metrics to evaluate performance on the FUNSD dataset, which can be downloaded at \url{https://guillaumejaume.github.io/FUNSD/}.

\end{abstract}


\IEEEpeerreviewmaketitle

\section{Introduction}

Forms are a common way to collect data. They are used in various fields, from medical reports to administrative data collection.
We define form understanding (FoUn) as the task of automatically extracting and structuring written information in a form. FoUn is based on text detection and recognition. Firstly, it analyzes the spatial layout and written information to identify the questions, answers, and headers present in the form. Secondly, it aims to understand how the extracted entities are interlinked. Here, we introduce the FUNSD dataset, a dataset for form understanding in noisy scanned documents. To the best of our knowledge, FUNSD is the first publicly available dataset that addresses FoUn task. The FUNSD dataset contains 199 fully annotated forms that vary widely with regard to their structure and appearance. The forms come from different fields, \textit{e.g.,} marketing, advertising, and scientific reports. They are all one-page forms rendered in a rasterized format with low resolution and corrupted by real noise. The forms were annotated in a bottom-up approach, allowing the FUNSD dataset to be used for various document-understanding tasks including text detection, text recognition, spatial layout understanding, and question-answer pair extraction. 

Extracting information from scanned documents is not a new task. For instance, previous work focused on digitizing the contents of documents into a machine-readable format using optical character recognition (OCR). See~\cite{IIN2017,TBA2016} for reviews of current OCR systems. Existing datasets include the ICDAR Robust Reading Competitions 2011, 2013, 2015, and 2017\footnote{\url{http://rrc.cvc.uab.es/?ch=1&com=introduction#}}. 
Another task of information extraction from documents is layout analysis that attempts to extract the content of a document and restore its structure by analyzing its spatial arrangement. Applications of layout analysis range from text and non-text separation to full text segmentation of complex layouts~\cite{CAP2018,EGO2017,MRK2003,Mar2013,YYA2017}.

An application closely related to FoUn is table understanding~\cite{AEA2018,FFC2017}. In this case, the goal is to retrieve the key-value pairs that map headers from a table to the value represented by a cell.  However, the tabular structure is rather rigid and is far from being as generic as the representation of forms.

Commercial solutions such as ABBYY\footnote{https://www.abbyy.com/}, Nuance\footnote{https://www.nuance.com/print-capture-and-pdf-solutions.html} or Datacap\footnote{https://www.ibm.com/ch-fr/marketplace/document-capture-and-imaging} allow information extraction from user-defined areas in specific pages of documents, including forms. This requires manual annotation of zones where an answer is expected to appear. However, these solutions do not scale well as the number of templates increases. On the contrary, the FUNSD dataset was created to build template-agnostic representations of forms. Moreover, FoUn goes beyond the aforementioned approaches and aims to extract structured information in a semantically meaningful way so that, for instance, it can be stored in a database, which in turn can be used for data analysis.

Our contributions can be summarized as follows:
\begin{itemize}
    \item We formalize form understanding as a series of defined tasks. From an image of a form, we define a pipeline to structure the textual content as a list of labeled semantic entities that are interlinked. 
    \item We provide access to the FUNSD dataset, a document understanding dataset for text detection, OCR, spatial layout analysis, and entity linking in noisy scanned forms. 
    \item We build a set of baselines that define the current state-of-the-art results for the FUNSD dataset.  
    \item We propose a set of metrics to evaluate the form understanding pipeline. 
\end{itemize}

\section{Dataset Description}
 
\subsection{A subset of the RVL-CDIP dataset}

To ensure that real data are used, featuring highly varying form structures and realistic noise, we used a subset of the RVL-CDIP dataset\footnote{https://www.cs.cmu.edu/~aharley/rvl-cdip/}~\cite{Harley2015}. The RVL-CDIP dataset is composed of $400,000$ grayscale images of various documents from the 1980s--1990s. Each image is labeled by its type, \textit{e.g.,} letter, email, magazine, form. The documents have a low resolution of around $100$~dpi. The images are also of low quality with various types of noise added by successive scanning and printing procedures. To build the FUNSD dataset, we manually checked the $25,000$ images from the form category. We discarded unreadable and similar forms, resulting in $3,200$ eligible documents, out of which we randomly sampled $199$ to annotate. Note that the RVL-CDIP dataset is a subset of the Truth Tobacco Industry Document\footnote{https://www.industrydocuments.ucsf.edu/tobacco/} (TTID), an archive collection of scientific research, marketing, and advertising documents of the largest US tobacco firms. The TTID archive aims to advance information retrieval research. 

\subsection{Annotation procedure}

The annotations used for text detection were performed by Figure8 mechanical turks\footnote{https://www.figure-eight.com/}. The remaining tasks were annotated using an annotation tool specifically designed for form understanding. The annotation tool is based on GuiZero\footnote{https://lawsie.github.io/guizero/}, a high-level library based on tkinter\footnote{http://tkinter.fdex.eu/}. 

\subsection{Dataset structure and format}

Each form is encoded in a JSON file. We represent a form as a list of semantic entities that are interlinked. A semantic entity represents a group of words that belong together from a semantic and spatial standpoint. Each semantic entity is described by a unique identifier, a label (\textit{i.e.,} question, answer, header or other), a bounding box, a list of links with other entities, and a list of words. Each word is represented by its textual content and its bounding box. All the bounding boxes are represented by their coordinates following the schema $\mbox{box} = [\mathbf{x}_{left}, \mathbf{y}_{top}, \mathbf{x}_{right}, \mathbf{y}_{bottom}]$. The links are directed and formatted as $[\mathbf{id}_{from}, \mathbf{id}_{to}]$, where $\mathbf{id}$ represents the semantic entity identifier.
The dataset statistics are shown in Table~\ref{tab:stat1}. Even with a limited number of annotated documents, we obtain a large number of word-level annotations ($>30$k) and entities ($\approx 10$k), making this dataset suitable for deep learning applications. 
The semantic entity class distribution is shown in Table~\ref{tab:stat2}. Naturally, the most common classes are questions and answers. 
\begin{table}[h]
    \caption{Dataset statistics.}
    \label{tab:stat1}
  \centering
  \begin{tabular}{l|c|c|c|c}
  \hline
    Split & Forms & Words & Entities & Relations \\
    \hline
    Training & $149$ & $22,512$ & $7,411$ & $4,236$ \\
    Testing  & $50$ & $8,973$ & $2,332$ & $1,076$ \\
    \hline
  \end{tabular}
\end{table}
\begin{table}[h]
    \caption{Class distribution of the semantic entities.}
    \label{tab:stat2}
  \centering
  \begin{tabular}{l|c|c|c|c|c}
  \hline
    Split & Header & Question & Answer & Other & Total \\
    \hline
    Training & $441$ & $3,266$ & $2,802$ & $902$ & $7,411$ \\
    Testing  & $122$ & $1,077$ & $821$ & $312$ & $2,332$ \\
    \hline
  \end{tabular}
\end{table}

\begin{figure}[h] 
    \centering
    \includegraphics[width=0.45\textwidth]{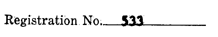}
    \caption{Screenshot of a form from the FUNSD dataset.}
    \label{fig:semantic_entity_example}
\end{figure}

An example of a ground-truth file is shown in Listing~\ref{list:json_format}. The corresponding sub-part of the original form is shown in Figure~\ref{fig:semantic_entity_example}. In this example, we have two semantic entities, \textit{``Registration No.''}, which is tagged as a question and \textit{``533''}, which is tagged as an answer. There is a link from the first semantic entity to the second one, resulting in a question--answer pair. 

\begin{lstlisting}[float,language=json,label=list:json_format,firstnumber=1,caption=Example of ground-truth format.]
{
    "form": [
        {
            "id": 0,
            "text": "Registration No.",
            "box": [94,169,191,186],
            "linking": [
                [0,1]
            ],
            "label": "question",
            "words": [
                {
                    "text": "Registration",
                    "box": [94,169,168,186]
                },
                {
                    "text": "No.",
                    "box": [170,169,191,183]
                }
            ]
        },
        {
            "id": 1,
            "text": "533",
            "box": [209,169,236,182],
            "label": "answer",
            "words": [
                {
                    "box": [209,169,236,182],
                    "text": "533"
                }
            ],
            "linking": [
                [0,1]
            ]
        }
    ]
}
\end{lstlisting}

\subsection{Limitations of the FUNSD dataset}

The main difficulty when building a dataset for form understanding applications is to ensure that the annotated corpus contains enough variability. Indeed, the visual representation of a form can vary drastically from one industry to another (\textit{e.g.,} a medical report vs a tax form). The range of variability comes mostly from the fact that there is no exact definition of \emph{what} a form is or \emph{how} we should represent it. In the FUNSD dataset, we attenuate this problem by selecting forms from different fields (marketing, science, advertisement, etc.). Nevertheless, we cannot ensure that we have captured enough examples to create a generic and generalizable form understanding application. 
Another limitation is that most of the textual content is machine-written. In many real-life scenarios, we expect to encounter handwritten text as well. Note that we still observe handwritten content in some of the forms, especially for signatures and dates. 

\section{Baselines and Metrics}

We present baseline results for text detection, text recognition, and form understanding on the FUNSD dataset. 

\subsection{Text detection}

We test text detection at the word level. State-of-the-art algorithms follow a data-driven approach. Usually, CNN-feature maps are extracted using a deep neural network. The network then predicts heat maps that represent the probability of whether a given pixel is part of a text and combines these heat maps with bounding box proposals~\cite{Zhou2017,Tian2018,Huang2018,Dai2017a}.

Text detection on the FUNSD dataset was tested with four baselines: Tesseract~\cite{Smith2007}, EAST~\cite{Zhou2017}\footnote{\url{https://github.com/argman/EAST}}, Google Vision API\footnote{\url{https://cloud.google.com/vision/docs/detecting-fulltext}}, and a Faster R-CNN architecture~\cite{Ren2017}. Tesseract, EAST, and Google Vision are tested without retraining on the FUNSD training set. As EAST and Google Vision output their predictions as quadrangles (\textit{i.e.,} four vertices that define a polygon), and the FUNSD dataset is annotated with rectangles, we transform each quadrangle into a rectangle by constructing the smallest rectangle that contains the four quadrangle vertices. The Faster R-CNN baseline is based on a PyTorch implementation\footnote{\url{https://github.com/facebookresearch/maskrcnn-benchmark}} that was retrained specifically for this task. We used a network pretrained on ImageNet with a ResNet-101 architecture~\cite{He2016}. We used anchors of sizes $(16, 32, 64, 128, 256)$, strides of $(4, 8, 16, 32, 64)$, and aspect ratios of $(0.5, 1.0, 2.0, 4.0, 8.0)$. During testing, we allow a maximum of $500$ object detections and select all objects with confidence detection $\> 0.5$. The learning rate was set to $10^{-3}$ with a weight decay of $0.0001$. The batch size was set to $1$ and the maximum number of epochs to $10$ with early stopping. For each approach, we compute the precision, recall, and F1 score  of the FUNSD test set at $\mbox{IoU} = 0.5$.
Results are shown in Table~\ref{tab:text_detection}. 
\begin{table}[h]
    \caption{Results for word-level text detection. Precision and recall expressed in \%.}
    \label{tab:text_detection}
  \centering
  \begin{tabular}{l|c|c|c}
  \hline
    Method & Precision & Recall & F1-score \\
    \hline
    Tesseract & $45.4$ & $68.0$ & $0.54$ \\
    EAST & $51.6$ & $84.0$ & $0.64$\\
    Google Vision & $\mathbf{79.8}$ & $62.0$ & $0.69$ \\
    Faster R-CNN & $70.4$ & $\mathbf{84.8}$ & $\mathbf{0.76}$ \\
    \hline
  \end{tabular}
\end{table}

Faster R-CNN baseline yields the best overall performance (\textit{i.e.,} highest F1-score). This observation is expected as we are specifically retraining the network for the task. Note that Google Vision also performs well, even without being retrained on the task, thus showing its generalization power. 

\subsection{Text recognition with optical character recognition}

OCR engines are usually based on appearance features to obtain a character-level prediction that is coupled with a sequence modeling network (\textit{e.g.,} LSTM, GRU) to extract the words~\cite{Borisyuk2018}. Modern engines that also support handwritten text recognition usually use a connectionist temporal classification (CTC) loss to cope with the alignment problem~\cite{Borisyuk2018}. Note that some novel architectures perform text detection and recognition in an end-to-end manner~\cite{Sui2018}.
 
We evaluate the relevance of the OCR output by computing the Levenshtein similarity between the predicted word $w_p$ and the ground-truth word $w_{gt}$:
%
\begin{align} \label{eq:lev_sim}
    S(w_p, w_{gt}) = 1 - \frac{L(w_p, w_{gt})}{\max(|w_p|, |w_{gt}|)}
\end{align}
where $L(w_p, w_{gt})$ is the Levenshtein distance between $w_p$ and $w_{gt}$ and $|.|$ denotes the number of characters in a word. The similarity is case sensitive and takes into account the recognition of checkboxes (often encountered in documents like forms). 
We evaluate two OCR engines for text recognition: Tesseract~\cite{Smith2007} and Google Vision. We evaluate OCR performance using two metrics, referred to as (1) text detection + OCR and (2) OCR.  In both cases, we compute the Levenshtein similarity between the correctly detected words and the ground truth (\textit{i.e.,} $\mbox{IoU}>0.5$). In the first case, we normalize by the total number of ground-truth words, whereas in the second case, we normalize by the number of \emph{identified} words. Note that no preprocessing is applied to the documents before feeding them to the OCR. 
\begin{table}[h]
    \caption{OCR results based on Levenshtein similarity. Results expressed in \%.}
    \label{tab:ocr}
  \centering
  \begin{tabular}{l|c|c}
  \hline
    Method & Text detection + OCR & OCR \\
    \hline
    Tesseract & $3.4$ & $7.3$  \\
    Google Vision  & $\mathbf{76.4}$ & $\mathbf{94.4}$ \\
    \hline
  \end{tabular}
\end{table}

Table \ref{tab:ocr} shows that Google Vision is a strong OCR baseline that captures  the textual content almost perfectly when the words are correctly identified ($\approx 95\%$). The Tesseract OCR engine performs poorly on the FUNSD dataset, which can be explained by the fact that the minimum quality of $300$~dpi needed by Tesseract is not met in the FUNSD dataset. 

\subsection{Form understanding}

We decompose the FoUn challenge into three tasks, namely word grouping, semantic-entity labeling, and entity linking.
\begin{itemize}
\item \textbf{Word grouping} is the task of aggregating words that belong to the same semantic entity. 
\item \textbf{Semantic entity labeling} is the task of assigning to each semantic entity a label from a set of four pre-defined categories: question, answer, header or other. 
\item \textbf{Entity linking} is the task of predicting the relations between semantic entities.
\end{itemize}

\begin{figure}[h] 
    \centering
    \includegraphics[width=0.5\textwidth]{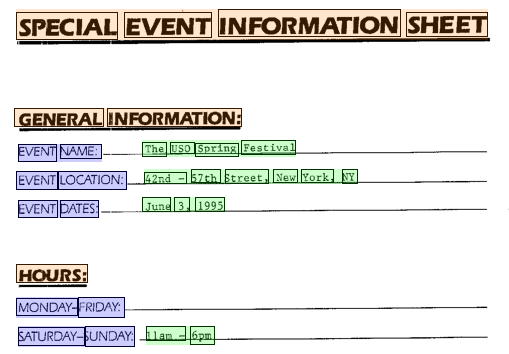}
    \caption{Example of word grouping and labeling. \textit{Questions} are represented in blue, \textit{headers} in orange, and \textit{answers} in green.}
    \label{fig:funsd_text}
\end{figure}

Figure~\ref{fig:funsd_text} illustrates this concept by showing the word grouping and labeling in a form of the FUNSD dataset. 

\subsubsection{Word grouping}

We tested the word grouping on two naive baselines based on textline extraction performed by Tesseract and Google Vision OCR engines. We propose that word grouping can be evaluated as a clustering problem, where words are the data points and clusters are the semantic entities. The optimal number of clusters is the number of semantic entities in the ground truth. All the words that were not recognized by the text detector (\textit{i.e.,} $\mbox{IoU}<0.5$), are assigned to a new artificial cluster. We propose using the adjusted rand index~\cite{Hubert1985} (ARI) as a metric. The ARI is based on the number of pairs correctly assigned to the same cluster adjusted to compensate for randomness. The results are presented in Table~\ref{tab:grouping}. As expected, the baselines perform poorly as they do not take into consideration the spatial layout and the textual content. We foresee a need for \emph{learned} algorithms for grouping the words in order to build more competitive algorithms. 

\begin{table}[h]
    \caption{Baseline results for word grouping. A value of $0$ corresponds to a random assignment and $1$ to a perfect clustering.}
    \label{tab:grouping}
  \centering
  \begin{tabular}{l|c}
  \hline
    Method & Word grouping \\
    \hline
    Tesseract & $0.20$   \\
    Google Vision & $\mathbf{0.41}$  \\
    \hline
  \end{tabular}
\end{table}

\subsubsection{Semantic entity labeling}

We propose using a simple learned neural baseline based on a multi-layer perceptron. We build input features for each semantic entity with 

\begin{itemize}
    \item semantic features extracted from the pretrained language model BERT~\cite{devlin2018bert}\footnote{\url{https://github.com/huggingface/pytorch-pretrained-BERT}},
    \item spatial features based on the bounding box coordinates of the semantic entity,
    \item meta features that encode the length of the sequence.
\end{itemize}

The resulting input feature dimension for each entity is $733$. Each semantic entity is then independently passed through an MLP with two hidden layers and $500$ units each with ReLU activation. The last layer is a softmax classifier to derive the class label.
Note that we test the algorithms by assuming that we know the optimal word grouping, word location, and textual content. In this way, we \emph{only} assess the specific task. Results are shown in Table~\ref{tab:labeling}. 

\begin{table}[h]
    \caption{Baseline results for the entity labeling and linking. Precision and recall expressed in \%.}
    \label{tab:labeling}
  \centering
  \begin{tabular}{l|c|c|c}
  \hline
    Task & Precision & Recall & F1-score \\
    \hline
    Entity labeling & $-$ & $-$ & $0.57$  \\
    Entity Linking & $2.1$ & $99.2$ & $0.04$  \\
    \hline
  \end{tabular}
\end{table}

\subsubsection{Entity linking}

For this step, we reuse the semantic entity input features built for the entity labeling task. We approach the entity-linking task as a binary classification task (\textit{i.e.,} to determine whether a link exists). We simply concatenate the feature representation of each semantic entity for all the possible pairs in the form. We then pass it through a MLP with two hidden layers and $500$ hidden units with ReLU activation. 

The metric we used verifies whether the predicted links exist among all the semantic entities correctly identified and labeled. We can then compute the precision, recall, and F1-score. Note that not all the semantic entities have relations with other semantic entities (e.g., a sentence describing the page number of the form or an unanswered question). Results are presented in Table~\ref{tab:labeling}. Stronger baselines should include the relational side of semantic entities that can naturally be represented as a graph. 


\section{Conclusion}

We introduced FUNSD, a new dataset for form understanding in noisy scanned documents along with a set of simple baselines and metrics to evaluate form understanding. We believe that this work can serve as a starting point for progress in the field of document understanding. Approaches to address form-understanding challenges include the development of an end-to-end deep learning pipeline that, given a set of words, jointly learns how to group them, assign a label, and build relations between them. 

\bibliography{library}
\bibliographystyle{IEEEtran}






\end{document}